\documentclass[12pt]{article}

\usepackage{axodraw}

\input{epsf}
\def\slash#1{\ooalign{$\hfil/\hfil$\crcr$#1$}} 

\def\lra{\leftrightarrow}

\newcommand\as{\alpha_{\mathrm{S}}}
\newcommand\gs{g_{\mathrm{S}}}

\def\ep{\epsilon}

\def\beq{\begin{equation}}
\def\eeq{\end{equation}}
\def\beeq{\begin{eqnarray}}
\def\eeeq{\end{eqnarray}}
\def\cm{{\cal M}}
\def\bom#1{{\mbox{\boldmath $#1$}}}
\def\to{\rightarrow}

\newcommand{\la}{\langle}
\newcommand{\ra}{\rangle}

\def\nn{\nonumber}

\def\ID{1 \kern -.45 em 1}
\def\RS{{\scriptscriptstyle\rm R\!.S\!.}}
\def\cdr{{\scriptscriptstyle\rm C\!.D\!.R\!.}}
\def\dimr{{\scriptscriptstyle\rm D\!.R\!.}}

\def\sp{{\bom {Sp}}}

\def\ket#1{|{#1}\ra}
\def\bra#1{\la{#1}|}

\def\ubar{{\overline u}}

\def\vep{{\varepsilon}}

\def\Eq#1{Eq.~({\ref{#1}})}

\def\ta#1{#1\hspace{-.42em}/\hspace{-.07em}} 
\def\ln#1{\mathrm{ln}\left(#1\right)}

\begin{document}

\begin{titlepage}
\renewcommand{\thefootnote}{\fnsymbol{footnote}}
\begin{flushright}
     CERN--TH/2003-206\\  hep-ph/0312067
     \end{flushright}
\par \vspace{10mm}

\begin{center}
{\Large \bf
The triple collinear limit of\\[1ex]
one-loop QCD amplitudes~\footnote{Work supported in part 
by EC 5th Framework Programme under contract number HPMF-CT-2000-00989.}
}
\end{center}
\par \vspace{2mm}
\begin{center}
{\bf Stefano Catani}~$^{(a)}$\footnote{E-mail: stefano.catani@fi.infn.it}, 
{\bf Daniel de Florian}~$^{(b)}$\footnote{E-mail: deflo@df.uba.ar}, and 
{\bf Germ\'an Rodrigo}~$^{(c,d)}$\footnote{E-mail: german.rodrigo@cern.ch}

\vspace{5mm}

${}^{(a)}$INFN, Sezione di Firenze and
Dipartimento di Fisica, Universit\`a
di Firenze,\\
I-50019 Sesto Fiorentino, 
Florence, Italy \\
\vspace*{2mm}
${}^{(b)}$Departamento de F\'\i sica, FCEYN, Universidad de Buenos Aires, \\
(1428) Pabell\'on 1 Ciudad Universitaria, 
Capital Federal, Argentina \\
\vspace*{2mm}
${}^{(c)}$Theory Division, CERN, CH-1211 Geneva 23, Switzerland \\
\vspace*{2mm}
${}^{(d)}$Instituto de F\'{\i}sica Corpuscular, E-46071 Valencia, Spain \\

\vspace{5mm}

\end{center}

\par \vspace{2mm}
\begin{center} {\large \bf Abstract} \end{center}
\begin{quote}
\pretolerance 10000

We consider the singular behaviour of one-loop QCD matrix elements
when several external partons become simultaneously parallel.
We present a new factorization formula that describes the singular collinear
behaviour directly in colour space. The collinear singularities are embodied
in process-independent splitting matrices that depend on the momenta, 
flavours, spins
and colours of the collinear partons. We give the general structure of the
infrared and ultraviolet divergences of the one-loop
splitting matrices. We also present explicit one-loop results for the triple
collinear splitting, $q \to q {\bar Q} Q$, of a quark and a quark--antiquark
pair of different flavours. The one-loop triple collinear splitting
is one of the ingredients that can be used to compute the evolution of parton
distributions at the next-to-next-to-leading order in QCD perturbation theory.

\end{quote}

\vspace*{\fill}
\begin{flushleft}
     CERN--TH/2003-206 \\ December 2003 
\end{flushleft}
\end{titlepage}

\renewcommand{\thefootnote}{\fnsymbol{footnote}}

The high precision of experiments at past (LEP), present (HERA, Tevatron)
and future (LHC, $e^+e^-$ linear colliders) particle colliders demands 
a corresponding precision in theoretical predictions. 
As for perturbative QCD predictions, this means calculations beyond 
the next-to-leading order (NLO) in the strong coupling $\as$. 
Recent years have witnessed much progress in this field
(see Ref.~\cite{2looprev} and references therein).
In particular, a great deal of work 
has been devoted to study the 
properties of QCD scattering amplitudes in the infrared (soft and collinear)
region \cite{Berends:1988zn}--\cite{Kosower:2002su}. 

The understanding of the infrared singular behaviour of multiparton 
QCD amplitudes is a
prerequisite for the evaluation of infrared-finite cross sections (and, 
more generally, infrared- and collinear-safe QCD observables) at the 
next-to-next-to-leading order (NNLO) in perturbation theory
\cite{2looprev}.
The information on the infrared properties of the amplitudes have also been
exploited to compute large (logarithmically enhanced) perturbative terms
and to resum them to all perturbative orders \cite{resex}.
The investigation of these properties is also valuable for improving the physics
content of Monte Carlo event generators
(see e.g. Ref.~\cite{mcws}). In addition, the results of these studies
prove to be useful beyond the strict QCD context, 
since they can provide 
hints on the structure of highly symmetric gauge theories at
infinite orders in the perturbative expansion
(see e.g. Ref.~\cite{Anastasiou:2003kj}).

In this paper we consider the collinear limit of multiparton QCD amplitudes
at one-loop order. We present a general factorization formula, which is valid
directly in colour space,
and discuss some properties of its infrared-divergent contributions.
These results apply to the multiple collinear limit of an arbitrary number of
QCD partons. As an example of application beyond the double collinear limit,
we present the result of the explicit evaluation 
of a triple collinear configuration of three quarks.
Besides its interest within the general framework outlined above,
our study of the one-loop triple collinear limit has specific
relevance to the NNLO calculation of the
Altarelli--Parisi kernels that control the scale evolution
of parton densities and fragmentation functions \cite{book}. 
This formidable NNLO computation is being completed
\cite{nnloap} by using traditional methods. 
Kosower and Uwer \cite{Kosower:2003np} have proposed
to exploit
collinear factorization at the amplitude level as an alternative method 
to perform the NNLO calculation of the Altarelli--Parisi kernels.
To this purpose, the one-loop triple collinear splitting
is one of the necessary ingredients. Two other ingredients are 
the tree-level quadruple collinear splitting
\cite{DelDuca:1999ha} and the two-loop double collinear splitting.
A detailed discussion of the multiple collinear limit and of
the results presented in this letter
will appear in a forthcoming paper \cite{inprep}.

We consider a generic scattering process involving 
final-state QCD partons (massless quarks and gluons) with momenta 
$p_1, p_2, \dots$ Non-QCD partons $(\gamma^*, Z^0, W^\pm, \dots)$ 
are always understood. The corresponding 
matrix element is denoted by
\beq
\label{meldef}
\cm^{c_1,c_2,\dots;s_1,s_2,\dots}_{a_1,a_2,\dots}(p_1,p_2,\dots) \;\;,
\eeq
where $\{c_1,c_2,\dots\}$, $\{s_1,s_2,\dots\}$ and $\{a_1,a_2,\dots\}$ are
respectively colour, spin and flavour indices. 
To take into account the colour and spin structures, 
we use the notation of Refs.~\cite{Catani:1998bh,Catani:1996vz}.
We introduce an orthonormal basis
$\{ \ket{c_1,c_2,\dots} \otimes \ket{s_1,s_2,\dots} \}$
in colour + spin space, in such a way that the matrix element
in Eq.~(\ref{meldef}) 
can be written as
\beq
\label{cmmdef}
\cm^{c_1,c_2,\dots;s_1,s_2,\dots}_{a_1,a_2,\dots}(p_1,p_2,\dots)
\equiv
\Bigl( \bra{c_1,c_2,\dots} \otimes \bra{s_1,s_2,\dots} \Bigr) \;
\ket{\cm_{a_1,a_2,\dots}(p_1,p_2,\dots)} \;.
\eeq
Thus $\ket{\cm_{a_1,a_2,\dots}(p_1,p_2,\dots)}$
is a vector in colour + spin space.
It is important to specify that the matrix elements $\cm(p_1,p_2,\dots)$
that we are considering are physical ones. Their external legs
are on shell $(p_i^2=0)$ and have physical spin polarizations.

The matrix element $\cm(p_1,p_2,\dots)$ can be evaluated in QCD perturbation
theory as a power series expansion (i.e. loop expansion) in the strong 
coupling $\as$ ($\as=\gs^2/(4\pi)$).
Throughout the paper 
we are mainly interested in 
the expansion up to one-loop order. We write
\beq
\label{loopex}
\cm = 
\left( \gs \right)^q
\; \left[ \; \cm^{(0)} 
+ \frac{\as}{2\pi} \;\cm^{(1)}  
+ {\cal O}(\as^2) \right] \;\;,
\eeq
where the overall power $q$ is integer ($q=0,1,2,3,\dots$).
In the evaluation of the one-loop amplitude $\cm^{(1)}$, one encounters 
ultraviolet and infrared singularities that have to be properly regularized.
We use dimensional regularization in $d=4- 2\ep$ space-time dimensions.
The dimensional-regularization scale is denoted by $\mu$.

The multiple collinear limit of the matrix element in Eq.~(\ref{meldef})
is approached when the momenta $p_1, \dots, p_m$ of $m$ partons become parallel.
This implies that all the particle subenergies
$s_{ij}=(p_i+p_j)^2$, with $i,j=1,\dots,m$, are of the same order and vanish
simultaneously. We thus introduce a pair of back-to-back
light-like (${\widetilde P}^2=0, n^2=0$)
momenta ${\widetilde P}^\nu$ and $n^\nu$, and we write
\beq
\label{vectorn}
(p_1 + \dots + p_m)^\nu = {\widetilde P}^\nu  
+ \frac{s_{1\dots m} \; n^\nu}{2 \, n \cdot {\widetilde P}} \;, \quad
s_{1\dots m} = (p_1 + \dots + p_m)^2 \;,
\eeq
where $s_{1\dots m}$ is the total invariant mass of the system of collinear
partons. In the collinear limit, the vector ${\widetilde P}^\nu$
denotes the collinear direction, and we have
$p_i^\nu \to z_i {\widetilde P}^\nu$, where the longitudinal-momentum 
fractions $z_i$ are
\beq
z_i = \frac{n \cdot p_i}{n \cdot {\widetilde P}} 
\eeq
and fulfil the constraint $\sum_{i=1}^m z_i =1$.
To be definite, in the rest of the paper we limit ourselves to
explicitly considering the collinear limit in the time-like region
($s_{ij} > 0, \;  1>z_i > 0$).

In the limit when the $m$ parton momenta $p_1, \dots, p_m$ become 
simultaneously parallel, the matrix element $\cm(p_1,\dots,p_m,p_{m+1},\dots)$
becomes singular. At the tree level, the dominant singular behaviour is
$\cm^{(0)}(p_1,\dots,p_m,p_{m+1},\dots)\sim (1/{\sqrt s})^{m-1}$,
where $s$ generically denotes
a two-particle subenergy $s_{ij}$, or a three-particle subenergy $s_{ijk}$,
and so forth. At one-loop order, this singular behaviour is simply
modified by scaling violation,
$\cm^{(1)}(p_1,\dots,p_m,p_{m+1},\dots)\sim (1/{\sqrt s})^{m-1} 
(s/\mu^2)^{-\ep}$.
The dominant singular behaviour can be captured by universal 
(process-independent) 
factorization formulae \cite{Bern:zx}--\cite{Kosower:2002su}.
At the matrix element level, the factorization formulae are usually presented
upon decomposition in colour subamplitudes.
Nonetheless, collinear factorization is valid directly in colour space
\cite{inprep}, and we present the factorization formulae in this more general 
form.

The colour-space factorization formulae for the multiple
collinear limit of the tree-level and one-loop amplitudes $\cm^{(0)}$ 
and $\cm^{(1)}$ are:
\beq
\label{facttree}
\ket{\cm_{a_1,\dots,a_m,a_{m+1},\dots}^{(0)}(p_1,\dots,p_m,p_{m+1},\dots)}
\simeq \sp_{a_1 \dots a_m}^{(0)}(p_1,\dots,p_m) 
\;\;\ket{\cm_{a,a_{m+1},\dots}^{(0)}({\widetilde P},p_{m+1},\dots)}
\;\;,
\eeq
\beeq
\label{factone}
\ket{\cm_{a_1,\dots,a_m,a_{m+1},\dots}^{(1)}(p_1,\dots,p_m,p_{m+1}\dots)}
&\!\!\simeq&\! \sp_{a_1 \dots a_m}^{(1)}(p_1,\dots,p_m) 
\;\;\ket{\cm_{a,a_{m+1}\dots}^{(0)}({\widetilde P},p_{m+1},\dots)} \nn \\
&\!\!+&\! \sp_{a_1 \dots a_m}^{(0)}(p_1,\dots,p_m) 
\;\;\ket{\cm_{a,a_{m+1},\dots}^{(1)}({\widetilde P},p_{m+1},\dots)}
\;. \nn \\
\eeeq
These factorization formulae are valid in any number $d=4-2\ep$ 
of space-time dimensions or, equivalently, at any order in the $\ep$ 
expansion around $d=4$.
The only approximation involved
on the right-hand side amounts to neglecting terms that are less singular
in the multiple collinear limit.
Equations (\ref{facttree}) and (\ref{factone})
relate the original
matrix element (on the left-hand side)
with $m+k$ partons (where $k$ is arbitrary) to a matrix element (on the
right-hand side) with $1+k$ partons. 
The latter is obtained from the former
by replacing the $m$ collinear partons 
with a single parent parton, whose momentum is ${\widetilde P}$ 
and whose 
flavour $a$ is determined by flavour conservation in the splitting process
$a \to a_1+ \dots+ a_m$.

The process dependence of the factorization formulae is entirely embodied
in the matrix elements. The tree-level and one-loop factors 
$\sp^{(0)}$ and $\sp^{(1)}$, which
encode the singular behaviour in the multiple collinear limit,
are universal (process-independent). They depend on the momenta and quantum
numbers (flavour, spin, colour) of the $m$ partons that arise from the collinear
splitting. According to the notation in Eq.~(\ref{cmmdef}),
$\sp$ is a matrix in colour+spin space,
and we name it the {\em splitting matrix}. Making the dependence on the
colour and spin indices explicit, we have
\beq
{\rm Sp}_{a_1 \dots a_m}^{(c_1,\dots,c_m;s_1,\dots,s_m) (c_a,s_a)} =
\Bigl( \bra{c_1,\dots,c_m} \otimes \bra{s_1,\dots,s_m} \Bigr)
\sp_{a_1 \dots a_m} 
\Bigl( \ket{c_a} \otimes \ket{s_a} \Bigr) \;\;,
\eeq
so that in Eqs.~(\ref{facttree}) and (\ref{factone})
$\sp(p_1,\dots,p_m)$ acts onto the colour and 
spin indices ($\{c_1,\dots,c_m;$ $s_1,\dots,s_m\}$) 
of the $m$ collinear partons 
on the left and onto the colour and spin indices ($\{c_a;s_a\}$) 
of the parent parton on the right.

The essential difference between Eqs.~(\ref{facttree}) and (\ref{factone})
and the known collinear-factorization formulae discussed in the literature
\cite{Bern:zx,Kosower:1999xi}
regards the role of the colour quantum number. The derivation of the
tree-level factorization formula (\ref{facttree}) in colour space
is quite straightforward \cite{Catani:1998nv}. Its one-loop extension,
Eq.~(\ref{factone}), is less straightforward and, in particular, it exploits
colour coherence of QCD radiation \cite{inprep}. The factorization formulae of
Refs.~\cite{Bern:zx,Kosower:1999xi} are written in terms of colour 
subamplitudes and process-independent splitting amplitudes denoted by
${\rm Split}(p_1,\dots,p_m)$. Generically speaking, the splitting amplitudes
${\rm Split}$ can be regarded as colour-stripped components of the splitting
matrices $\sp$. 
For example, in the case of the tree-level collinear splitting
$g \to q_1+{\bar q}_2$, the splitting matrix is
\beq
\label{s0qbarq}
{\rm Sp}_{q_1 {\bar q}_2}^{(0) \,(\beta_1,\beta_2) (c)}(p_1,p_2) = 
\mu^\ep \;t^{c}_{\beta_1 \beta_2} \; \frac{1}{s_{12}} \; \ubar(p_1)\, 
{\slash \vep}^*({\widetilde P}) \,v(p_2) \;\;,
\eeq
where $t^c$ $(c=1,\dots,N_c^2-1)$ are the $SU(N_c)$ colour matrices in the
fundamental representation (we use the normalization ${\rm Tr}( t^a t^b )=
T_R \,\delta^{ab} , \,T_R=1/2$),
$u, v$ are customary Dirac spinors, $\vep$ is the physical
polarization vector of the parent gluon, and the corresponding
spin indices are understood. 
The splitting amplitude
${\rm Split}_{q_1 {\bar q}_2}^{(0)}$ is obtained from Eq.~(\ref{s0qbarq})
by simply removing the colour factor $t^{c}_{\beta_1 \beta_2}$.
Analogous proportionality relations apply to any tree-level and one-loop
splitting process $a \to a_1+a_2$. When the splitting involves
$m \geq 3$ collinear partons, splitting matrices and splitting amplitudes
are not simply proportional. Nonetheless, the two formulations 
of collinear factorization are related 
by gauge invariance and colour
algebra, and each formulation has its own practical advantages.

Several general properties of the one-loop splitting amplitudes
${\rm Split}^{(1)}(p_1,p_2)$ and, consequently, of $\sp^{(1)}(p_1,p_2)$ 
were discussed in 
Refs.~\cite{Bern:zx,Bern:1998sc,Kosower:1999rx}. Here, we present an additional
general property of $\sp^{(1)}(p_1,\dots,p_m)$. 

When computed in
$d=4-2\ep$ space-time dimensions, the one-loop amplitude $\cm^{(1)}$ and, 
hence, $\sp^{(1)}$ have ultraviolet and infrared divergences that
show up as $\ep$-poles in the expansion around  the point $\ep=0$.
To make the divergent behaviour explicit, we write the 
one-loop contribution to the splitting matrix as
\beq
\label{sp1df}
\sp^{(1)}(p_1,\dots,p_m) = \sp^{(1) \,{\rm div.}}(p_1,\dots,p_m) + 
\sp^{(1) \,{\rm fin.}}(p_1,\dots,p_m) \;\;,
\eeq
where $\sp^{(1) \,{\rm div.}}$ behaves as $1/\ep^2$ and
$\sp^{(1) \,{\rm fin.}}$ is finite when $\ep \to 0$. We also recall that
different regularization schemes (RS) can actually be implemented
within the dimensional-regularization
prescription. Two customary RS are the conventional 
dimensional-regular\-ization (CDR) scheme \cite{cdruv} and the
dimensional-reduction (DR) scheme \cite{dred}.
The RS dependence of the tree-level and one-loop 
splitting matrices $\sp^{(0)}$ and $\sp^{(1)}$ starts at 
${\cal O}(\ep^0)$ and at ${\cal O}(1/\ep)$, respectively.
In Eq.~(\ref{sp1df}), we define $\sp^{(1) \,{\rm fin.}}$ in such a way that 
it is RS-independent at 
${\cal O}(\ep^0)$, though it is still RS-dependent at higher orders in the
$\ep$-expansion. Therefore, $\sp^{(1) \,{\rm div.}}$ contains the full
RS-dependence of $\sp^{(1)}$ at ${\cal O}(1/\ep^2)$, ${\cal O}(1/\ep)$
and ${\cal O}(\ep^0)$.

The divergent part of the one-loop splitting matrix can be evaluated along the
lines of Refs.~\cite{Catani:1998bh,Catani:1996vz,Catani:2000ef}, and
the result can be expressed in terms of a process-independent factorization 
formula. We obtain \cite{inprep}
\beeq
\!\!\! \sp^{(1) \,{\rm div.}}(p_1,\dots,p_m) &=& 
\frac{\Gamma(1+\epsilon) \,
\Gamma^2(1-\epsilon)}{\left(4\pi\right)^{-\epsilon}\Gamma(1-2\epsilon)}
\; \frac{1}{2}
\left\{ \frac{1}{\ep^2}
\sum_{i,j=1 (i \neq j)}^m \;{\bom T}_i \cdot {\bom T}_j
\left( \frac{-s_{ij} -i0}{\mu^2}\right)^{-\ep} \right. \nn \\
&&\left. + 
\left( \frac{-s_{1 \dots m} -i0}{\mu^2}\right)^{-\ep}
\left[ \frac{1}{\ep^2}
\sum_{i,j=1}^m \;{\bom T}_i \cdot {\bom T}_j
\;\left( 2 - \left( z_i \right)^{-\ep} -\left( z_j \right)^{-\ep} \right)
\right. \right.  \nn \\
\label{sp1div}
&&\left. \left. - \frac{1}{\ep} 
\left( \sum_{i=1}^m \left( \gamma_i - \ep {\tilde \gamma}_i^{\RS} \right)
- \left( \gamma_a - \ep {\tilde \gamma}_a^{\RS} 
\right) - \frac{m-1}{2} \left( \beta_0 - \ep {\tilde \beta}_0^{\RS}
\right) \right) \right]
\right\} \nn \\
&\times& \sp^{(0)}(p_1,\dots,p_m) \;\;,
\eeeq
where we have used the same notation as in Ref.~\cite{Catani:1998bh}.
The colour charge (matrix) of the collinear parton with momentum $p_i$
is denoted by ${\bom T}_i$, and colour conservation implies
$\sum_i {\bom T}_i \;\sp^{(0)} = \sp^{(0)} \;{\bom T}_a$ (${\bom T}_a$
is the colour charge of the parent parton in the collinear splitting).
The flavour coefficients $\gamma_i$ 
and $\beta_0$
are $\gamma_q=\gamma_{\bar q}=3C_F/2$
and $\gamma_g=\beta_0/2=(11C_A-2N_f)/6$. The flavour coefficients 
${\tilde \gamma}_i^{\RS}$ and ${\tilde \beta}_0^{\RS}$ are RS-dependent 
\cite{rsdep}. In particular,
${\tilde \gamma}_i^{\cdr}={\tilde \beta}_0^{\cdr}=0$, while 
${\tilde \gamma}_q^{\dimr}={\tilde \gamma}_{\bar q}^{\dimr}=C_F/2$
and ${\tilde \gamma}_g^{\dimr}={\tilde \beta}_0^{\dimr}/2=C_A/6$.

Note that, so far, we have not specified whether the one-loop
amplitudes $\cm^{(1)}$ and splitting matrices $\sp^{(1)}$
are renormalized or unrenormalized quantities. Since the renormalization 
procedure commutes with the collinear limit, the factorization formula
(\ref{factone}) equally applies to both renormalized and unrenormalized 
quantities. However, the divergent part $\sp^{(1) \,{\rm div.}}$ 
explicitly given in Eq.~(\ref{sp1div}) refers to the (charge) unrenormalized
splitting matrix (thus, $\as$ is the bare QCD coupling). 
In the curly bracket of Eq.~(\ref{sp1div}), 
the contribution proportional to $\beta_0 - \ep {\tilde \beta}_0^{\RS}$
is of ultraviolet origin, and it would disappear by working at the level of
renormalized matrix elements and splitting matrices.

It is straightforward to check that Eq.~(\ref{sp1div}) agrees
with the divergent behaviour of the known 
\cite{Bern:zx,Bern:1998sc,Kosower:1999rx}
one-loop splitting amplitudes for the double collinear limit $a \to a_1+a_2$.
The triple collinear splitting process explicitly considered below provides
us with a further check of the general result in Eq.~(\ref{sp1div}).

At the tree level,
the splitting amplitudes are known for all possible partonic 
channels in the double \cite{Berends:1987me,Mangano:1990by} and triple 
\cite{DelDuca:1999ha} collinear limits, and for pure gluonic splitting 
($g\to 4g$) also in the quadruple collinear case \cite{DelDuca:1999ha}. 
In the double collinear case, the square of the splitting amplitudes
gives the leading-order Altarelli--Parisi kernels \cite{Altarelli:1977zs}.
The square of the triple collinear splitting amplitudes was computed 
in Refs.~\cite{Campbell:1997hg,Catani:1998nv}, and checked in 
Ref.~\cite{DelDuca:1999ha}.

At the one-loop level, the splitting amplitudes for all partonic 
channels in the double collinear limit were computed  
in Refs.~\cite{Bern:1998sc,Kosower:1999rx} to all orders in $\ep$.
In the following we consider the triple collinear limit
and, more precisely, the collinear splitting process
$q \to q_1+\bar{Q}_2+Q_3$, where $q$ and $Q$ denote (massless)
quarks of different flavours.

We first recall the tree-level splitting matrix for the process
$q \to q(p_1)+\bar{Q}(p_2)+Q(p_3)$:
\beeq
\label{s0qqbarq}
{\rm Sp}_{q_1 {\bar Q}_2 Q_3}^{(0) \,(\beta_1,\beta_2,\beta_3)
(\beta)}(p_1,p_2,p_3) &=& 
\mu^{2\ep} \, \sum_c t^{c}_{\beta_3 \beta_2} t^{c}_{\beta_1 \beta}  \\
&\times&
\frac{1}{s_{123} s_{23}} 
\; \ubar(p_3) \, \gamma^\mu \, v(p_2)
\; \ubar(p_1) \, \gamma^\nu \, u({\widetilde P}) \; d_{\mu\nu}(p_2+p_3,n)~,
\nn
\eeeq
where $\beta_i \,(i=1,2,3)$ and $\beta$ are respectively the colour indices 
of the final-state fermions and of the parent quark, and
\beq
\label{gpolten}
d^{\mu \nu}(k,n)=-g^{\mu \nu} + \frac{k^{\mu} n^{\nu}+n^{\mu} 
k^{\nu}}{n\cdot k}
\eeq
is the physical polarization tensor of the gluon with momentum $k$
($n^\mu$ is the auxiliary light-like vector introduced in Eq.~(\ref{vectorn})).

The square of the splitting matrix $\sp_{a_1 \dots a_m}$,
summed over final-state colours and spins 
and averaged over colours and spins of the parent parton, defines the 
$m$-parton splitting function $\langle \hat{P}_{a_1 \cdots a_m} \rangle$, 
which is a generalization of the customary (i.e. with $m=2$)
Altarelli--Parisi splitting function. 
Fixing the normalization of the tree-level splitting function 
$\langle \hat{P}_{a_1 \cdots a_m}^{(0)} \rangle$ by
\beq
\label{p0def}
\langle \hat{P}_{a_1 \cdots a_m}^{(0)} \rangle = 
\left( \frac{s_{1 \dots m}}{2 \;\mu^{2\ep}} \right)^{m-1} \;
{\overline {| \sp_{a_1 \dots a_m}^{(0)} |^2}} \;\;,
\eeq
from \Eq{s0qqbarq}
we have
\beq
\label{p0res}
\langle \hat{P}^{(0)}_{q_1 \bar{Q}_2 Q_3} \rangle = 
\frac{1}{2} C_F T_R \frac{s_{123}}{s_{23}}
\bigg[ -\frac{t^2_{23,1}}{s_{23}s_{123}} + \frac{4z_1+(z_2-z_3)^2}{z_2+z_3}
+(1-2\epsilon \delta^{\RS}) 
\left( z_2+z_3-\frac{s_{23}}{s_{123}} \right) \bigg]~,
\eeq
where
\beq
t_{ij,k}\equiv 2 \frac{z_i \ s_{jk}-z_j \ s_{ik}}{z_i+z_j}+
\frac{z_i-z_j}{z_i+z_j} s_{ij}~,
\eeq
which agrees with the result obtained in 
Refs.~\cite{Campbell:1997hg,Catani:1998nv}. The parameter $\delta^{\RS}$
depends on the RS: $\delta^{\cdr}=1$ and $\delta^{\dimr}=0$.

To evaluate the one-loop splitting matrix $\sp_{a_1 \dots a_m}^{(1)}$
we use a process-independent method \cite{Catani:1998nv,Kosower:1999xi,inprep}.
In the case of the collinear splitting process $q \to q_1+\bar{Q}_2+Q_3$,
$\sp_{q_1 \bar{Q}_2 Q_3}^{(1)}$ receives a contribution from two different colour
structures:
\beeq
\label{s1qqbarq}
{\rm Sp}_{q_1 {\bar Q}_2 Q_3}^{(1) \,(\beta_1,\beta_2,\beta_3)
(\beta)}(p_1,p_2,p_3) =
\mu^{4\ep} \, \frac{8\pi^2}{s_{123}} &&\!\!\!\!\!\!\!\! \left\{ 
\sum_c t^{c}_{\beta_3 \beta_2} t^{c}_{\beta_1 \beta} 
\;{\cal S}(p_1,p_2,p_3) \right. \\
+&&\!\!\!\! \left.
\sum_{b,c} (t^b t^c+t^c t^b)_{\beta_3 \beta_2} (t^c t^b)_{\beta_1 \beta} \; 
\;{\cal A}(p_1,p_2,p_3) \right\} \;. \nn
\eeeq
The first term in the curly bracket has the same structure as the tree-level
contribution in Eq.~(\ref{s0qqbarq}). The colour structure of the second term
is a new one-loop (quantum) effect.

The one-loop correction $\hat{P}^{(1)}$ to the tree-level splitting function
is obtained by simply performing the replacement 
$| \sp^{(0)} |^2 \rightarrow (\sp^{(0)})^\dagger \, \sp^{(1)} 
+ (\sp^{(1)})^\dagger \, \sp^{(0)}$ on the right-hand side of
Eq.~(\ref{p0def}). From Eqs.~(\ref{s0qqbarq}) and (\ref{s1qqbarq}),
we obtain
\beq 
\label{p1qqbarq}
\langle \hat{P}_{q_1 \bar{Q}_2 Q_3} \rangle = 
\langle \hat{P}^{(0)}_{q_1 \bar{Q}_2 Q_3} \rangle +
\frac{\as}{2\pi} \left[ 
\; \langle \hat{P}^{(1) \,\mathrm{(sym.)}}_{q_1 \bar{Q}_2 Q_3} \rangle
+ \langle \hat{P}^{(1) \,\mathrm{(an.)}}_{q_1 \bar{Q}_2 Q_3} \rangle \;
\right] + {\cal O}(\as^2)~.
\eeq
Note that $\langle \hat{P}^{(0)}_{q_1 \bar{Q}_2 Q_3} \rangle$ is symmetric
with respect to the exchange  
$p_2 \lra p_3$ of the $\bar Q_2$ and $Q_3$ momenta (see Eq.~(\ref{p0res})).
Its one-loop correction has, instead, a symmetric component and an 
antisymmetric component, respectively denoted by 
$\la \hat{P}^{(1) \,\mathrm{(sym.)}}_{q_1 \bar{Q}_2 Q_3} \ra$ and
$\la \hat{P}^{(1) \,\mathrm{(an.)}}_{q_1 \bar{Q}_2 Q_3} \ra$.
The antisymmetric component is entirely produced by the one-loop
splitting amplitude ${\cal A}(p_1,p_2,p_3)$ on the right-hand side of
Eq.~(\ref{s1qqbarq}).

\begin{figure}
\begin{center}
\begin{picture}(300,120)(0,0)
\Text(60,0)[]{(a)}
\Text(240,0)[]{(b)}
\SetOffset(0,20)
\Vertex(50,60){1.5}
\Vertex(30,10){1.5}
\Vertex(65,10){1.5}
\Vertex(85,60){1.5}
\ArrowLine(0,10)(30,10)
\ArrowLine(30,10)(65,10)
\ArrowLine(65,10)(110,10)
\ArrowLine(50,60)(50,100)
\ArrowLine(110,60)(85,60)
\ArrowLine(85,60)(50,60)
\Gluon(30,10)(50,60){4}{5}
\Gluon(85,60)(65,10){4}{5}
\Text(0,0)[]{$p_{123}$}
\Text(90,0)[]{$p_1$}
\Text(45,0)[]{$p_1-q$}
\Text(120,60)[]{$p_2$}
\Text(60,100)[]{$p_3$}
\Text(85,25)[]{$q$}
\Text(5,45)[]{$q+p_2+p_3$}
\Text(75,70)[]{$-(q+p_2)$}
\SetOffset(180,20)
\Vertex(50,60){1.5}
\Vertex(30,10){1.5}
\Vertex(65,10){1.5}
\Vertex(85,60){1.5}
\ArrowLine(0,10)(30,10)
\ArrowLine(30,10)(65,10)
\ArrowLine(65,10)(110,10)
\ArrowLine(50,60)(50,100)
\ArrowLine(110,60)(85,60)
\ArrowLine(85,60)(50,60)
\Gluon(30,10)(85,60){4}{6}
\Gluon(50,60)(65,10){4}{5}
\Text(0,0)[]{$p_{123}$}
\Text(90,0)[]{$p_1$}
\Text(45,0)[]{$p_1-q$}
\Text(120,60)[]{$p_2$}
\Text(60,100)[]{$p_3$}
\Text(75,25)[]{$q$}
\Text(15,35)[]{$q+p_2+p_3$}
\Text(70,70)[]{$q+p_3$}
\end{picture}
\end{center}
\caption{One-loop diagrams contributing to the antisymmetric component 
of the splitting matrix $\sp_{q_1 \bar{Q}_2 Q_3}$.}
\label{fig:oneloop}
\end{figure}
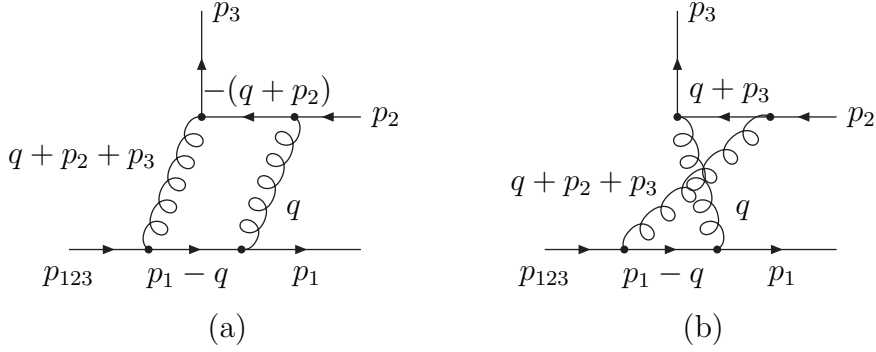

The expression of the splitting amplitude ${\cal A}(p_1,p_2,p_3)$
is sufficiently compact to be presented explicitly in this letter.
We have
\beeq
\label{aamp}
{\bom {\cal A}}(p_1,p_2,p_3)
&=& - \frac{1}{2} \, i \int \frac{d^d q}{(2\pi)^d} \; 
\ubar(p_3) \left[ \,\frac{ \, \gamma^\sigma (\ta{q}+\ta{p}_2) \gamma^\mu \,}
{(s_{2q}+i0)}  
- \frac{ \, \gamma^\mu (\ta{q}+\ta{p}_3) \gamma^\sigma \,}
{(s_{3q}+i0)} \, \right] v(p_2)\nn \\ &\times&
d_{\mu \nu}(q,n) \ d_{\sigma \rho}(q+p_2+p_3,n) \;
\frac{ \ubar(p_1) \, \gamma^\nu (\ta{p}_1 - \ta{q}) \gamma^\rho 
\, u({\widetilde P})}{(q^2+i0)(t_{1q}+i0)(s_{23q}+i0)} \;,
\end{eqnarray}
where 
\beq
t_{1q}=(p_1-q)^2~, \qquad s_{iq}=(p_i+q)^2~, \qquad s_{23q}=(p_2+p_3+q)^2~.
\eeq
The first and second contributions in the square bracket originate from 
the one-loop diagrams depicted in 
Figs.~\ref{fig:oneloop}a and \ref{fig:oneloop}b, respectively.

Note that the expression in Eq.~(\ref{aamp}) is valid in any RS.
It is also valid in any number $d=4-2\ep$ of space-time dimensions or,
equivalently, to all orders in $\ep$. To make the dependence on $\ep$ explicit,
we have to compute the $d$-dimensional integral over the loop momentum $q$.
This computation requires the evaluation of a set of basic one-loop
(scalar and tensor) integrals. The corresponding integrands 
involve, besides the customary Feynman propagators $1/(q^2+i0)$, additional
propagators of the type $1/(n\cdot q)$, which come from the
physical polarizations of the virtual gluons (see Eq.~(\ref{gpolten})).
Some of these integrals, 
which resemble those encountered in axial-gauge
calculations, were evaluated by Kosower and Uwer \cite{Kosower:1999rx}
in the context of their
calculation of the one-loop double collinear splitting $a \to a_1+a_2$.
More complicated 
integrals (higher-point functions) of this type are involved in 
triple collinear splitting processes. We have computed 
(to high orders in the $\ep$ expansion) all the basic one-loop integrals
\cite{inprep}
that appear in any triple collinear splitting $a \to a_1+a_2+a_3$.
Using these results, we have obtained explicit expressions up to
${\cal O}(\ep^0)$ of the splitting amplitude 
${\cal A}(p_1,p_2,p_3)$ in Eqs.~(\ref{s1qqbarq}),~(\ref{aamp})  
and of the corresponding splitting function 
$\la \hat{P}^{(1) \,\mathrm{(an.)}}_{q_1 \bar{Q}_2 Q_3} \ra$ in Eq.~(\ref{p1qqbarq}).
We limit ourselves to presenting the expression of 
$\la \hat{P}^{(1) \,\mathrm{(an.)}}_{q_1 \bar{Q}_2 Q_3} \ra$, since the 
expression of ${\cal A}(p_1,p_2,p_3)$ has a very similar structure.

Including terms up to ${\cal O}(\ep^0)$, our result for the 
antisymmetric component of the one-loop splitting function is
\begin{eqnarray}
\label{finaleq}
& &  \!\!\!\!\!\!\!\!\!\!\!\!\!\!
\langle \hat{P}^{(1) \,\mathrm{(an.)}}_{q_1 \bar{Q}_2 Q_3} \rangle =  
\frac{\Gamma(1+\epsilon) \;
\Gamma^2(1-\epsilon)}{\left(4\pi\right)^{-\epsilon}\Gamma(1-2\epsilon)}
\; C_F \, T_R \, \frac{N_c^2 -4}{4N_c} \;
\left(\frac{-s_{123}-i 0}{\mu^2}\right)^{-\epsilon}
\nn \\ 
&\times&
\left( 
\left\{
\frac{1}{\epsilon^2}
\left[
\left( \left(\frac{s_{23}}{s_{123}}\right)^{-\epsilon} + 1 \right)
\left( \left(\frac{z_2}{z_2+z_3}\right)^{-\epsilon}
-\left(\frac{z_3}{z_2+z_3}\right)^{-\epsilon}
+ \left(\frac{s_{13}}{s_{123}}\right)^{-\epsilon}
- \left(\frac{s_{12}}{s_{123}}\right)^{-\epsilon}
\right) 
\right. \right. \right.
 \nn \\ 
&+& \left. 
\left(\frac{s_{12}}{s_{123}}\right)^{-\epsilon}
\left(\frac{z_3}{z_2+z_3}\right)^{-\epsilon}
- \left(\frac{s_{13}}{s_{123}}\right)^{-\epsilon}
\left(\frac{z_2}{z_2+z_3}\right)^{-\epsilon}
 \right]
\Bigg\} \la \hat{P}^{(0)}_{q_1 \bar{Q}_2 Q_3} \ra/(C_F T_R)
\nn \\
&+& 
\Bigg\{ \left( 
\hat{a} - \frac{s_{13}}{s_{123}-s_{13}} \ \hat{b} \right) \ln{\frac{s_{13}}{s_{123}}} 
+ \left( 
\frac{s_{23}}{s_{123}-s_{23}} \ \hat{a} -  \hat{b} \right) 
\ln{\frac{s_{23}}{s_{123}}} \nn \\
&+&\frac{s_{123} \ s_{12}}{s_{12}+s_{13}-z_1 \ s_{123}}
\left( \frac{z_2 \ \hat{a} - z_1 \ \hat{b}}{s_{12} (1-z_1)} + 
\frac{\hat{a}+\hat{b}}{s_{123}} \right) \ln{\frac{s_{12} \ z_2}{s_{13} \ z_3}} 
\ln{\frac{s_{23}}{s_{123} (z_2+z_3)}}  \nn \\
&+& 
\frac{(2s_{12}+s_{23}) \  \hat{a} + (s_{12}-s_{13}) \ \hat{b}}{s_{12}}
\nn \\ &\times& \! \left. \left.
\left[\ln{\frac{s_{13}}{s_{123}}} \ln{\frac{s_{23}}{s_{123}}}
+ \mathrm{Li}_2\left(1-\frac{s_{13}}{s_{123}}\right)
+ \mathrm{Li}_2\left(1-\frac{s_{23}}{s_{123}}\right) 
- \frac{\pi^2}{6} \right] - (2 \leftrightarrow 3) \right\} 
+ {\cal O} (\epsilon) \right) \nn \\
&+& {\rm complex \;\;conjugate} \, , \!\!
\end{eqnarray}
where
\beeq
\hat{a} &=& \frac{s_{123}}{s_{23}} \Bigg( \frac{z_1 \ s_{13}}{s_{123}}  
+\frac{z_1 (z_1 \ s_{23}-z_2 \ s_{13}-z_3 \ s_{12})}{2(z_2+z_3) \ s_{12}}
+\frac{z_1 \ s_{23}+z_2 \ s_{13}-z_3 \ s_{12}}{2 s_{123}} \Bigg)~, 
\\
\hat{b} &=& \frac{s_{123}}{s_{23}} \Bigg( \frac{z_2 \ s_{23}}{s_{123}}  
+\frac{z_2 (z_1 \ s_{23}-z_2 \ s_{13}+z_3 \ s_{12})}{2(z_2+z_3) \ s_{12}}
+\frac{z_1 \ s_{23}+z_2 \ s_{13}-z_3 \ s_{12}}{2 s_{123}} \Bigg)~. 
\eeeq

It is quite straightforward to check that the divergent part of the one-loop
splitting function in Eq.~(\ref{finaleq})
fully agrees with the result obtained by using the general formula in 
Eq.~(\ref{sp1div}). Note that the double poles $1/\ep^2$ cancel in
Eq.~(\ref{finaleq}), so that the most divergent terms in 
$\la \hat{P}^{(1) \,\mathrm{(an.)}}_{q_1 \bar{Q}_2 Q_3} \ra$ are single poles 
$1/\ep$. 
These single poles originate from the infrared region of the loop integral in
Eq.~(\ref{aamp}). More precisely, they arise when the loop momentum $q$ is soft
$(q \to 0)$ but not collinear to any of the external momenta.
Note also that, up to terms of ${\cal O}(\ep^0)$, the full 
RS dependence of $\la \hat{P}^{(1) \,\mathrm{(an.)}}_{q_1 \bar{Q}_2 Q_3} \ra$
is embodied in the corresponding dependence (see Eq.~(\ref{p0res}))
of the tree-level term 
$\la \hat{P}^{(0)}_{q_1 \bar{Q}_2 Q_3} \ra$ on the right-hand side of 
Eq.~(\ref{finaleq}).
The absence of additional RS-dependent terms (such as those proportional to
${\tilde \gamma}_i^{\RS}$ and ${\tilde \beta}_0^{\RS}$ in Eq.~(\ref{sp1div}))
at ${\cal O}(\ep^0)$ is 
related to the absence of single poles $1/\ep$
of ultraviolet and collinear origin, as discussed in the 
second paper of Ref.~\cite{rsdep}.

An additional check of our result can be performed 
by considering the triple collinear limit of known 
one-loop matrix elements, such as the matrix element for the process
$e^+e^- \to \bar{q} q \bar{Q} Q$ \cite{Bern:1996ka,Glover:1996eh}.
We have evaluated the one-loop splitting amplitude 
(\ref{aamp}) in a spin basis of definite $\pm 1$ helicities of the 
quarks and antiquarks. We have then compared the result with that
obtained by directly performing the corresponding collinear limit of the 
one-loop helicity amplitudes explicitly presented in Ref.~\cite{Bern:1996ka}
up to ${\cal O}(\ep^0)$ (note that the expressions in Ref.~\cite{Bern:1996ka}
refer to the DR scheme).
We have found complete agreement.

The splitting function $\hat{P}_{a_1 \cdots a_m}$ controls the singular
behaviour in the multiple collinear limit. Moreover, it
can have additional singularities
(i.e. terms that are not integrable in $d=4$ dimensions) in some 
subregions of the collinear phase-space. 
The tree-level splitting function $\la \hat{P}^{(0)}_{q_1 \bar{Q}_2 Q_3} \ra$
in Eq.~(\ref{p0res}) is indeed singular when the momenta $p_2$ and $p_3$
are parallel (i.e. when their
relative angle is much smaller than the emission angle of $p_1$)
and when they are simultaneously soft.
The singularity when $p_2$ and $p_3$ are parallel is, instead, absent 
(to any order in the $\ep$ expansion \cite{inprep}, 
and not only at ${\cal O} (\epsilon^0)$)
in the antisymmetric part, 
$\la \hat{P}^{(1) \,\mathrm{(an.)}}_{q_1 \bar{Q}_2 Q_3} \ra$,
of the one-loop splitting function. The singularity when $p_2$ and $p_3$
are simultaneously soft is still present in 
$\la \hat{P}^{(1) \,\mathrm{(an.)}}_{q_1 \bar{Q}_2 Q_3} \ra$, though its effect
vanishes after integration over the angles of $p_2$ and $p_3$
because of the antisymmetry with respect to the exchange 
$p_2 \leftrightarrow p_3$. The expression in Eq.~(\ref{finaleq}) shows that
$\la \hat{P}^{(1) \,\mathrm{(an.)}}_{q_1 \bar{Q}_2 Q_3} \ra$ has no other singularities
in any subregions of the phase-space.

For the sake of brevity, we have limited ourselves, in this letter,
to presenting a few explicit results for the one-loop triple collinear splitting.
These results have mainly an illustrative purpose.
The method and the tools (in particular, the one-loop integrals)
used to obtain them are sufficient and can be applied straightforwardly  
to evaluate the one-loop splitting matrix of any splitting process
$a \to a_1+a_2+a_3$.

\noindent{\bf Acknowledgements.} 
We wish to thank Werner Vogelsang for collaboration at an early stage of this
work. 
DdF thanks Fundaci\'on Antorchas and Conicet for partial support.
GR thanks Generalitat Valenciana under grants CTIDIB/2002/24 and GRUPOS03/013, 
and MCyT under grant FPA-2001-3031 for partial support.

\end{document}